\documentclass[reprint,aps,prb,twocolumn,superscriptaddress,showpacs]{revtex4}
\usepackage{amssymb}
\usepackage{upgreek}
\usepackage{graphicx}
\usepackage{mathrsfs}
\usepackage{textcomp}
\usepackage[utf8]{inputenc}
\usepackage{color}
\usepackage{siunitx}
\usepackage{url}

\usepackage{float}
\usepackage[paperwidth=210mm,paperheight=297mm,centering,hmargin=2cm,vmargin=2cm]{geometry}
\usepackage{lipsum}
\usepackage{circuitikz}
\usepackage{fancyhdr}
\usepackage{mathtools}
\usepackage{bm}
\usepackage[normalem]{ulem}
\usepackage{amsmath}

\begin{document}

\title{Transient magnetic domain wall AC dynamics by means of MOKE microscopy  }

\author{P. Domenichini}
\affiliation{Universidad de Buenos Aires, FCEyN, Departamento de F\'isica.  Buenos Aires, Argentina.}
\affiliation{CONICET-Universidad de Buenos Aires, IFIBA, Buenos Aires, Argentina.}
\author{C. P. Quinteros}
\affiliation{Instituto de Nanociencia y Nanotecnología, CNEA--CONICET, Centro Atómico Bariloche, (R8402AGP) San Carlos de Bariloche, Río Negro, Argentina.}
\affiliation{Zernike Institute for Advanced Materials, University of Groningen, 9747 AG Groningen, Netherlands.}
\author{M. Granada}
\affiliation{Instituto de Nanociencia y Nanotecnología, CNEA--CONICET, Centro Atómico Bariloche, (R8402AGP) San Carlos de Bariloche, Río Negro, Argentina.}
\author{S. Collin}
\affiliation{Unité Mixte de Physique, CNRS, Thales, Univ. Paris-Sud, Université Paris-Saclay, Palaiseau 91767, France.}
\author{J.-M. George}
\affiliation{Unité Mixte de Physique, CNRS, Thales, Univ. Paris-Sud, Université Paris-Saclay, Palaiseau 91767, France.}
\author{J. Curiale}
\affiliation{Instituto de Nanociencia y Nanotecnología, CNEA--CONICET, Centro Atómico Bariloche, (R8402AGP) San Carlos de Bariloche, Río Negro, Argentina.}
\affiliation{Instituto Balseiro, Universidad Nacional de Cuyo--CNEA, Av. E. Bustillo 9500, (R8402AGP) S. C. de Bariloche, Río Negro, Argentina.}
\author{S. Bustingorry}
\affiliation{Instituto de Nanociencia y Nanotecnología, CNEA--CONICET, Centro Atómico Bariloche, (R8402AGP) San Carlos de Bariloche, Río Negro, Argentina.}
\author{M. G. Capeluto}
\affiliation{Universidad de Buenos Aires, FCEyN, Departamento de F\'isica.  Buenos Aires, Argentina.}
\affiliation{CONICET-Universidad de Buenos Aires, IFIBA, Buenos Aires, Argentina.}
\author{G. Pasquini}
\affiliation{Universidad de Buenos Aires, FCEyN, Departamento de F\'isica.  Buenos Aires, Argentina.}
\affiliation{CONICET-Universidad de Buenos Aires, IFIBA, Buenos Aires, Argentina.}

\date{\today }


\begin{abstract}

The domain wall response under constant external magnetic fields reveals a complex behavior where sample disorder plays a key role. Furthermore, the response to alternating magnetic fields has only been explored in limited cases and analyzed in terms of the constant field solution. Here we unveil phenomena in the evolution of magnetic domain walls under the application of alternating magnetic fields within the creep regime, well beyond a small fluctuation limit of the domain wall position. Magnetic field pulses were applied in ultra-thin ferromagnetic films with perpendicular anisotropy, and the resulting domain wall evolution was characterized by polar magneto-optical Kerr effect microscopy.  Whereas the DC characterization is well predicted by the elastic interface model, striking unexpected features are observed under the application of alternating square pulses: magneto-optical images show that after a transient number of cycles, domain walls evolve toward strongly distorted shapes concomitantly with a modification of domain area.  The morphology of domain walls is characterized with a roughness exponent when possible and contrasted with alternative observables which result to be more suitable for the characterization of this transient evolution. The final stationary convergence as well as the underlying physics is discussed.

\end{abstract}

\maketitle

\section{Introduction} 
\label{sec:Introduction}

The dynamics and morphology of magnetic domain walls (DWs) play a key role on determining the spatial and temporal characteristics of magnetization reversal, whose control is imperative to develop new magnetization based devices~\cite{Allwood2005, Hayashi2008, Parkin2008, Stamps2014, Hellman2017, Sander2018}. Beside practical applications, from the basic point of view, the understanding of DW dynamics is also relevant on a wider context. DWs can be described within the general class of disordered elastic systems, which includes a vast range of systems such as ferroelectric domain walls \cite{Tybell2002, Paruch2013, ferroDW}, reaction fronts in disordered flows~\cite{Atis2015}, contact lines in wetting \cite{wetting}, epitaxially grown surfaces \cite{epitaxial}, crack propagation \cite{crack_propagation}, or active cell migration~\cite{Chepizhko2016}, as well as periodic systems such as vortex lattices in type II superconductors \cite{Blatter} or colloidal systems \cite{Reichhardt2016}, among others. 
Although the microscopic equations behind these systems are completely different, under some reliable assumptions all of them can be described as elastic manifolds lying in a disordered landscape \cite{Nattermann90, Kolton2006, Giammarchi2009, Agoritsas2012}. In those systems, the competition between elastic and pinning forces, together with thermal fluctuations, leads to a rich and complex dynamics. Remarkably, a thermally activated motion over effective drive-dependent energy barriers, described by universal laws, holds below the depinning transition \cite{Blatter, Lemerle98, Metaxas2007, Jeudy2016}.

In this context, the relationship that the morphology and spatial correlation lengths have with the associated dynamic regimes involves a very rich and interesting physics. Such a connection between dynamics and morphology is illustrated by the link between the universal creep exponent and the equilibrium roughness exponent. The creep exponent $\mu$ characterizes how the effective energy barriers depend on the external drive (magnetic field $H$ for the specific case we shall study), $\Delta E \sim H^{-\mu}$, which then controls the velocity-field characteristics below the depinning transition, $\ln v \sim -H^{-\mu}$ ~\cite{Nattermann90, Chauve2000}. The roughness exponent $\zeta$ describes how the fluctuations $B(r)$ of the position of the interface grow as a function of the length scale, $B(r) \sim r^{2 \zeta}$. The equilibrium roughness exponent thus describes the morphology of the interface at zero drive.
The creep exponent for a $d$-dimensional system can be expressed in terms of the roughness exponent as $\mu = (d+2 \zeta - 2)/(2 - \zeta)$~\cite{Nattermann90, Chauve2000}. Therefore, in a one-dimensional interface, the theoretically found equilibrium roughness exponent $\zeta = 2/3$~\cite{Huse85, Kardar1985} implies a creep exponent $\mu = 1/4$. This value of the creep exponent has been corroborated in many experiments \cite{Lemerle98, Metaxas2007, Ferre2013, Savero-Torres2018}. However, the experimental determination of the roughness exponent $\zeta$ is challenging and, while it can be computed in different ways~\cite{epitaxial, Lemerle98, Paruch2005, Metaxas2007, Bauer2005, Moon2013, Savero-Torres2018}, there is a wide spread in the reported values \cite{Paruch2005, Bauer2005, Moon2013, papercabrugosidad}. Furthermore, besides the equilibrium roughness exponent, the morphology can be described with different roughness exponents at different length scales~\cite{Kolton2006, Kolton2009, Ferrero2013}, which prompt to a careful interpretation of the experimentally found values~\cite{Grassi2018}.

Although a large amount of work during the last decades has been devoted to understanding this very rich phenomenology, there are still many unanswered questions. In particular, the description based on the analogy between magnetic domain walls and driven elastic interfaces considers several approximations;  the (quasi) equilibrium condition, elasticity in a small fluctuations limit (neglegting plasticity and multivalued interfaces), and the assumption of absence of strong pinning (or spatially correlated defects) in the sample, which are not obviously fulfilled. Several recent works highlight that some features can not be explained without taking into account some of these considerations \cite{papercabrugosidad, Caballero2018}.

In this framework, the AC dynamics of driven DWs has not been extensively explored up to now. After the seminal theoretical works of Natterman and coworkers \cite{Lyuksyutov1999, Nattermann2001, Glatz2003, Nattermann2004}, experiments carried out mainly by AC susceptibility measurements corroborated some of their predictions\cite{Kleemann2007}. The theoretical works\cite{Lyuksyutov1999, Nattermann2001, Glatz2003, Nattermann2004} generically study the response of DWs to AC fields and in particular predict different magnetic hysteresis loops as a function of the amplitude and frequency of an applied oscillating magnetic field, $H(t) = H_0 \sin \omega t$. It is predicted that at low frequencies $\omega < \gamma H_p/L$ and for amplitudes such that $H_0 > H_\omega$ the AC behavior is well described by applying the dynamic equation holding in the DC case. Here $\gamma$ is the mobility, $H_p$ is the depinning field, $L$ is a typical system size and $H_\omega$ is a frequency dependent field such that below $H_\omega$ the probability to overcome the creep energy barrier for the smallest DW segment is negligible and there is no macroscopic motion of the DW. The resulting solution only depends on the involved DC dynamic regimes and the typical distance $L$ between domain nuclei that determines the magnetic saturation~\cite{Nattermann2001}. In this picture, the application of an alternating periodic magnetic field with null DC component, would produce a periodic DW oscillation around the initial condition. Although the response is expected to be non-linear and hysteretic, the magnetic domain is expected to remain unchanged after applying an integer number of AC field cycles.
On the other hand, numerical calculations made in the context of the kinetic Ising model predict a dynamic phase transition as a function of the AC frequency \cite{Tome1990, DPT}, corroborated by magneto-optical Kerr effect magnetization measurements \cite{Robb2008, berger2013}: above a critical frequency a non-zero average magnetization develops in a transient number of cycles. 
In other  complex systems as superconducting vortices or colloidal assemblies, it has been shown that the AC dynamics displays very particular characteristics not directly translatable from DC dynamic regimes and may produce an evolution and a reorganization of the systems \cite{Reichhardt2016, Kawamura2017, Daroca2011, Marziali2015}. 

In the present work we show that DWs driven by moderate AC magnetic fields in the creep regime unveil new phenomena not directly associated with DC dynamics. By magneto-optically imaging the DW evolution under the application of alternating magnetic fields, in different ultra-thin ferromagnetic films with perpendicular magnetic anisotropy (PMA), we observe that after a transient number of cycles magnetic domains that are initially circular evolve toward strongly distorted domains, with irregular shape. This strong deformation of the DW morphology goes far beyond the small fluctuation limit \cite{Nattermann90, Giammarchi2009} and is concomitant with a reduction of domain area. We characterize the morphology of DWs with a roughness exponent when possible and we show that alternative observables take into account large scale deformations, are more suitable for the characterization of this transient evolution. The final stationary convergence as well as the underlying physics is discussed.

The paper is organized as follows: In Sec. \ref{sec:Methods} the experimental methods are described, the DC and AC protocols used to study different characteristics of the magnetic domains dynamics are described and the observables to quantify the measurements are defined. In Sec. \ref{sec:DC} we present the DC characterization of the samples used in this article. In Sec. \ref{sec:AC} we  show the main experimental results obtained with the AC protocol. Finally, Sec. \ref{sec:Conclusions} is devoted to a discussion and the main conclussions of the present work.

\section{Methods}\label{sec:Methods}

\subsection{\small Experimental techniques}

Magneto-optical imaging experiments have been performed in two kinds of ultra-thin ferromagnetic films with PMA. On one hand, Pt/Co/Pt magnetic monolayers were grown by DC magnetron sputtering on naturally oxidized (001) Si substrates at room temperature, as detailed in Ref.~\onlinecite{Cintia2018}. On the other hand, Pt/[Co/Ni]$_4$/Al multilayers were grown on oxidized Si-SiO$_2$ substrates by DC magnetron sputtering as described in Ref.~\onlinecite{Rojas2016}.  
Samples were characterized by DC out-of-plane magnetization measurements displaying the sharp square hysteresis loops characteristic of PMA. Both the Pt/Co/Pt monolayers \cite{Je2013, Hrabec2014, Lavrijsen2015, Wells2017} and the Co/Ni multilayers \cite{Rojas2016} samples present a sizeable Dzyaloshinski-Moriya (DMI) interaction with interfacial origin, which for instance can be evidenced using in-plane magnetic fields.
Results presented in this work correspond to the following samples: sample S1 is a Pt(8 nm)/Co(1 nm)/Pt(4 nm) containing one magnetic Co layer and sample S2 is a Pt(6 nm)/[Co(0.2 nm)/Ni(0.6 nm)]$_4$/Al(6 nm) containing a magnetic CoNi multilayer (the number in parenthesis indicates the thickness of each layer).

Magneto-optical images were obtained at room temperature, with a home-made polar magneto-optical Kerr effect (PMOKE) microscope in the Köhler configuration, using as light source a LED with a central wavelength of 650 nm. Two polarizers were included in the excitation and collection optical paths. $10\times$ amplified images were obtained with a 12 bits CCD  camera, with $0.39 \, \mu \mathrm{m/pixel}$ spatial resolution. Specially designed Helmholtz coils allowed us to apply square magnetic field pulses with amplitude $H$ up to 700 Oe and duration $\tau> 1\,\mathrm{ms}$, in the direction normal to the sample, with homogeneity $\Delta H/H < 0.04$ in a $0.126\,\mathrm{mm}^2$ area. 

\begin{figure}[t!]
  \centering
    \includegraphics[width= 6.5 cm ]{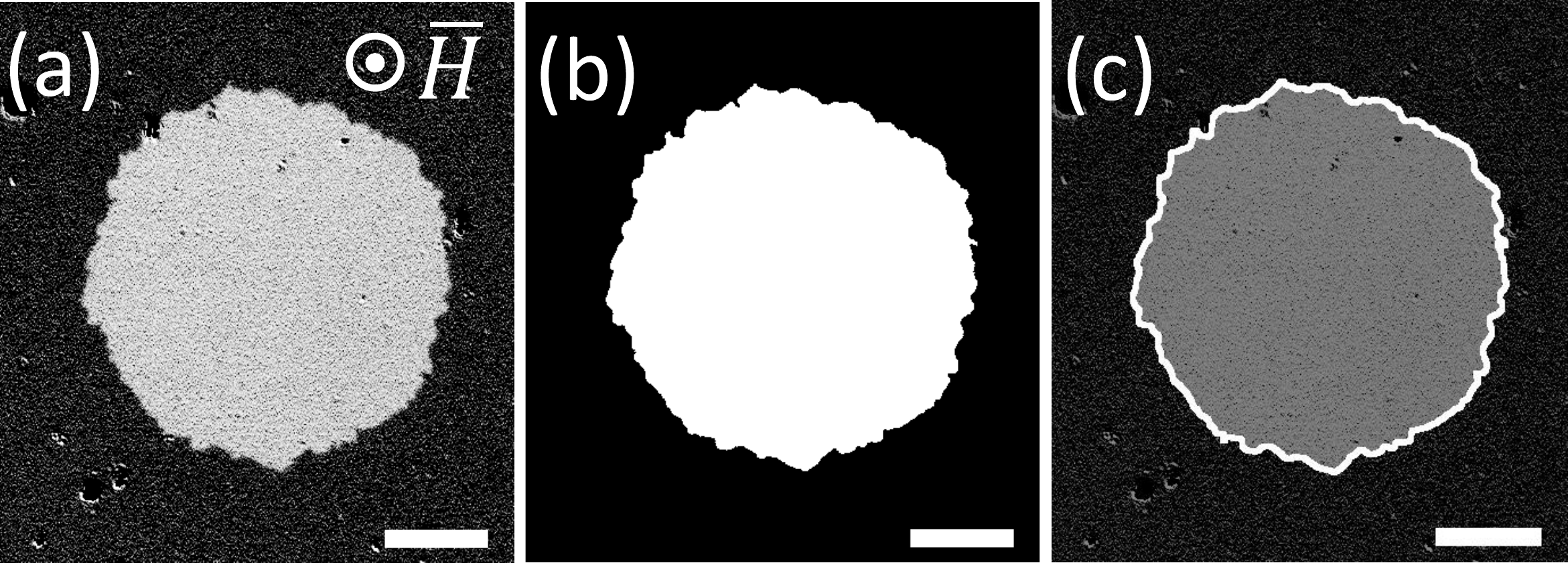}
    \caption{Example of image processing. (a) Image of a magnetic domain obtained after background subtraction (saturated sample). (b) After applying a threshold averaging filter a binary image is constructed, that allows to obtain the DW contour. (c) Superposition of the resulting DW profile, indicated as a white line, and the original image (panel a). The scale bar is $40 \, \mu \mathrm{m}$ width.}
    \label{fig:dominios}
\end{figure}

Figure \ref{fig:dominios}(a) shows a typical image of a magnetic domain after subtracting background (fully saturated sample). The contrast is further increased and the noise is cleaned using threshold averaging filter, that consists of applying a non-linear mapping function followed by a moving average and a threshold filter. The non-linear mapping function consists of $I(x,y)= \tanh(f(x,y)/I_{max})$, where $f(x,y)$ is the intensity of the image in each pixel after subtracting the background and $I_{max}$ is the maximum intensity of the image. The moving average filter is then used to smooth the noise~\cite{DIP-MAfilter}.  Finally, we applied a threshold filter to binarize the images as it is observed in Fig. \ref{fig:dominios}(b). As it is shown in Fig. \ref{fig:dominios}(c) by the superposition of the domain wall profile taken from the binarized image and the subtracted background image, there is not significant loss of information in the DW during this process.

\subsection{\small DC and AC protocols}

\begin{figure}[ht!]
  \centering
    \includegraphics[width=\linewidth]{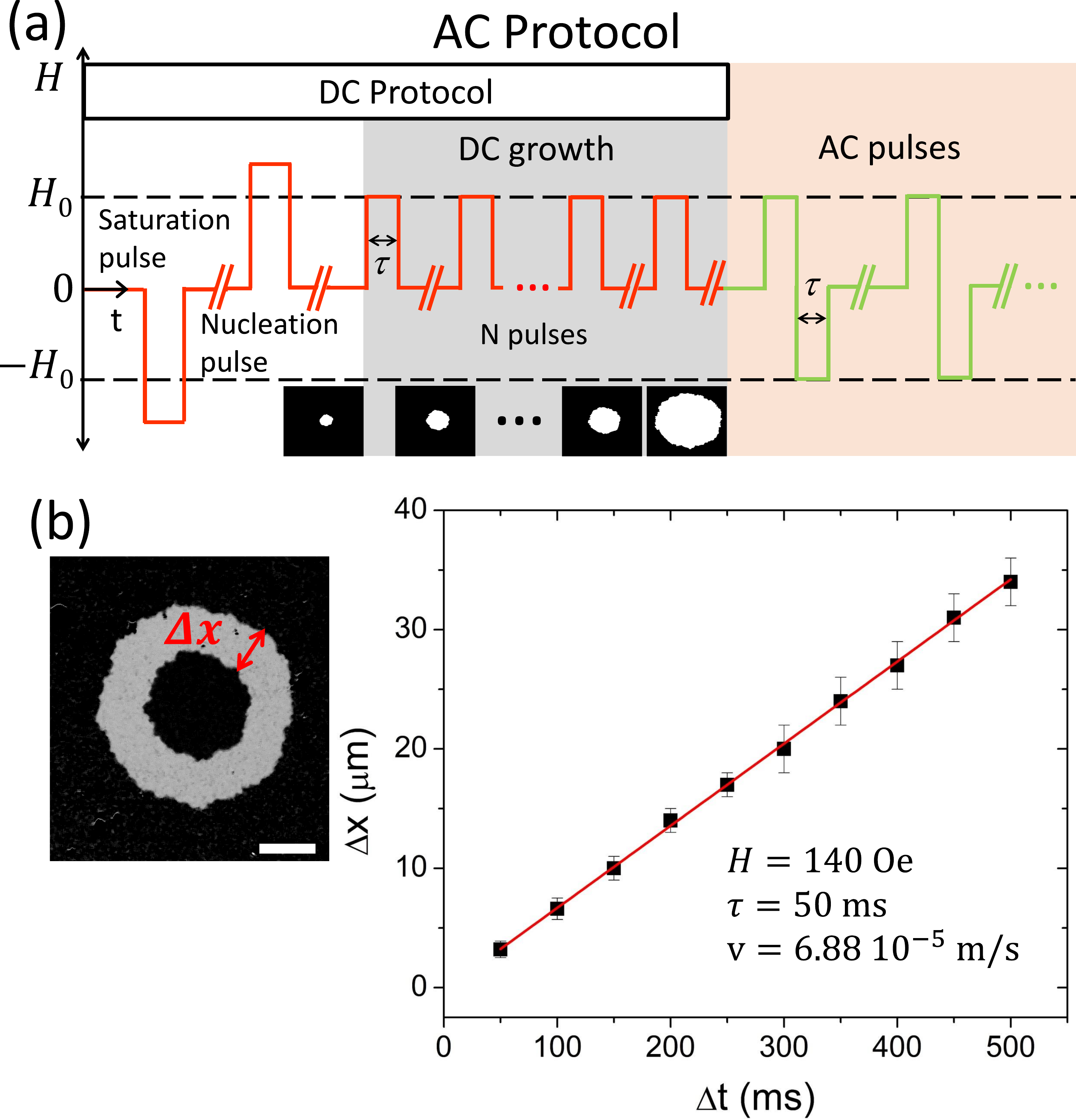}
    \caption{(Color online) Protocols to probe and characterize the DC and AC dynamics.  (a) Temporal evolution of the applied field $H$. The initial part corresponds to the DC protocol: After nucleation, the domains are grown by applying magnetic field pulses with the same polarity, $H > 0$. After DC growth, the AC magnetic field pulses of alternating polarity are applied, with null average magnetic field after application of an integer number of AC cycles. In both cases, images are obtained at $H = 0$, between (or after) pulses. (b) Example of the procedure followed to determine the DW mean velocity. The left panel shows the displacement $\Delta x$ of the DW after $N$ DC pulses of duration $\tau$ each. The scale bar is $40 \, \mu \mathrm{m}$. The velocity is obtained from the linear fit of $\Delta x$ against the total time $\Delta t = N \tau$.}
    \label{fig:protocolo}
\end{figure}

As in typical PMOKE experiments, due to the DW velocity scales involved, DW dynamics was characterized with a quasi-static technique in which the structure of the domains remains stable during image acquisition. Therefore, images of magnetic domains are taken in zero field after applying magnetic field square pulses that expand (or reduce) the domains (Fig. \ref{fig:protocolo}).

To measure the mean DW velocity, a sequence of pulses were applied [DC protocol in Fig. \ref{fig:protocolo}(a)], successive images were subtracted from one another and the mean displacement was measured (see Fig. \ref{fig:protocolo}(b), left panel). The time interval $\tau$ was chosen in such a way that the domains grow in small steps without reaching the magnetic saturation of the sample. In Fig. \ref{fig:protocolo}(b) the displacement after $N$ DC pulses is plotted against the total time duration $\Delta t = N \tau$. The velocity is then determined from the slope of a linear fit of the displacement data plotted against the total time. From now on, in agreement with the terminology used in the literature, we will refer to the procedure sketched under “DC Protocol” in Fig. \ref{fig:protocolo}(a) as a way to probe the “DC dynamics”.
On the other hand, to probe the “AC dynamics”, an already DC grown domain is shaken by applying several AC squared wave pulses, where the applied magnetic field alternates polarity and the average applied field is zero [Fig. \ref{fig:protocolo}(a)]. Amplitudes and times used in DC and AC protocols may be different. In all the cases, the DWs characteristics are measured from images taken under zero field.

\subsection{\small Correlations and observables}

\begin{figure}[ht!]
  \centering
    \includegraphics[width=\linewidth]{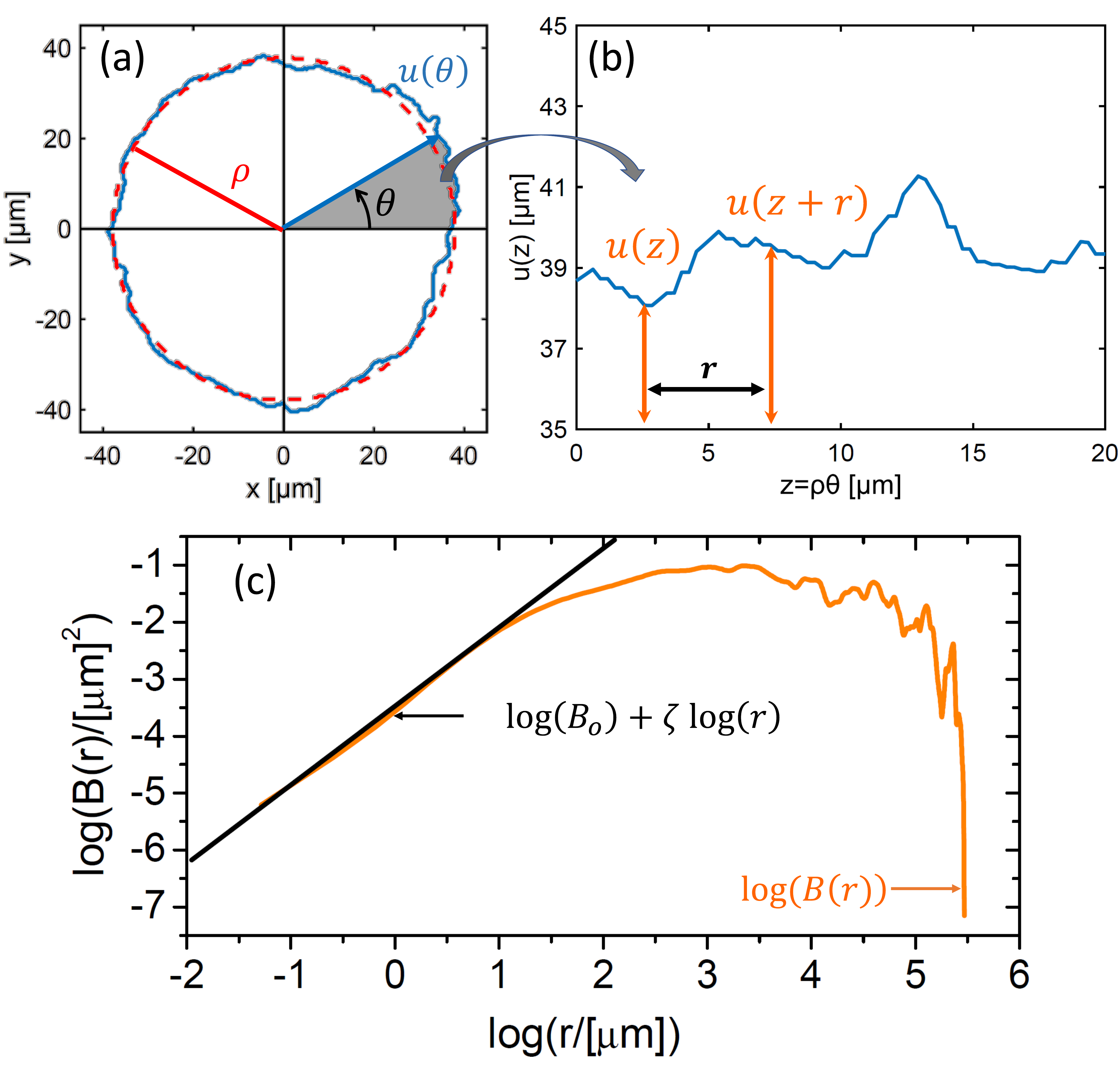}
    \caption{(Color online). Procedure used to compute the roughness function $B(r)$: A linear coordinate $z = \rho \theta$ is defined from a circle centered in the centroid of the domain with the same total magnetized area (panels a and b). $B(r)$ quantifies the correlation between $u(z)$ and $u(z + r)$ (panel b) as defined in Eq.~\eqref{FunBr}. The roughness exponent $\zeta$ and roughness amplitude $B_{0}$ are defined in Eq.~\eqref{rugosidad} and are obtained from the linear fit of $B(r)$ in log-log scale (panel c).}
    \label{fig:rugosidad}
\end{figure}

The simplest elastic model used to study response under AC fields assumes a flat interface in the presence of random weak defects~\cite{Nattermann2001, Glatz2003, Nattermann2004}. In the present work, we are particularly interested in describing the evolution of the full domain,  which of course is not flat. Since we are working with circular-type of domains, we alternatively solved this problem by transforming the angular position $\theta$ of each point on the DW in a linear coordinate, using a system centered in the centroid of the binarized magnetic domain. The linear position $z$ is then defined as $z = \rho \theta$, where $\rho$ is the radius of a circular domain with identical area than that of the observed domain. The procedure is sketched in Fig. \ref{fig:rugosidad}. The morphology of an interface is usually characterized by its roughness, an observable that quantifies the correlation between relative displacements of points in the interface a distance $r$ apart [see Fig. \ref{fig:rugosidad}(b)].   In this work we use as a measure of the interface fluctuations the roughness function
\begin{equation}\label{FunBr}
B(r)= \frac{1}{2\pi \rho }\int_{0}^{2\pi \rho} [u(z+r)-u(z)]^2 dz
\end{equation}
where $u(z)$ is the displacement in the radial direction (normal to the average DW position) and $2 \pi \rho$ is the effective interface length, as shown in Fig. \ref{fig:rugosidad}(a). Particular care has been taken to avoid DW regions pinned by strong defects, in order to compute $B(r)$.

As it was mention in the introduction, interface theory predicts relationships between different dynamic regimes and the corresponding roughness exponents $\zeta$~\cite{Kolton2006, Kolton2009, Ferrero2013}. In an elastic interface,  $\zeta$ characterizes the power-law growth of the roughness function $B(r)$,
\begin{equation}\label{rugosidad}
B(r)=B_{0} \, r^{2\zeta }.
\end{equation}
The exponent $\zeta$ quantifies correlations of transversal fluctuations in a single interface, whereas the roughness amplitude $B_{0}$ is related with the typical size of such fluctuations. From a linear fit of $\log B(r)$ vs. $\log r$ the roughness exponent $\zeta$ and roughness amplitude $B_{0}$ are obtained, as shown in Fig. \ref{fig:rugosidad}(c). 

The roughness function $B(r)$ serves to describe DWs morphology in a small fluctuation limit. Conversely, when fluctuations are large and the DW displacement is no longer univaluated, a different approach should be taken. Therefore, as a general tool permitting the characterization of domains evolution for arbitrarily large DW fluctuations, we further define a normalized correlation between two domain images $I_1 $ and $I_2 $ as
\begin{equation}\label{FunCorr}
{\cal C}(I_1, I_2)=\frac{max({\cal X}_{(I_1,I_2)})}{max({\cal X}_{(I_1,I_1)})}
\end{equation}
where the function ${\cal X}_{(I_1,I_2)}= I_1 \otimes	I_2$ is the 2-dimensional discrete cross-correlation function, that essentially performs the sliding scalar product between the binary images $I_1$ and $I_2$~\cite{DIP-MAfilter}. Specifically, for images of $M \times M$ pixels, where $I_1 (x_n, y_m)$ and $I_2 (x_n, y_m)$ are the values of the intensity at the pixel $(x_n, y_m)$, the value of the cross correlation in the position $(x_i,y_j)$ is computed as the inner product between the image $I_1$  and the image $I_2$ shifted in $(x_i,y_j)$ (also known as lags), this is
\begin{equation}\label{FunxCorr}
{\cal X}_{(I_1,I_2)}(x_i,y_j)=\sum_{n,m=1}^{M,M}I_1^* (x_n, y_m) I_2 (x_n + x_i, y_m + y_j),
\end{equation}
where $I_1^*$ is the complex conjugate of $I_1$ (that is equal to $I_1$ for a real image). As it will be shown in the following sections, ${\cal C}$ can be used to compare a domain after evolution with its initial condition or with an idealized perfect circular shaped domain. 

\section{DC dynamics characterization}\label{sec:DC}

\begin{figure}[ht!]
  \centering
    \includegraphics[width=\linewidth]{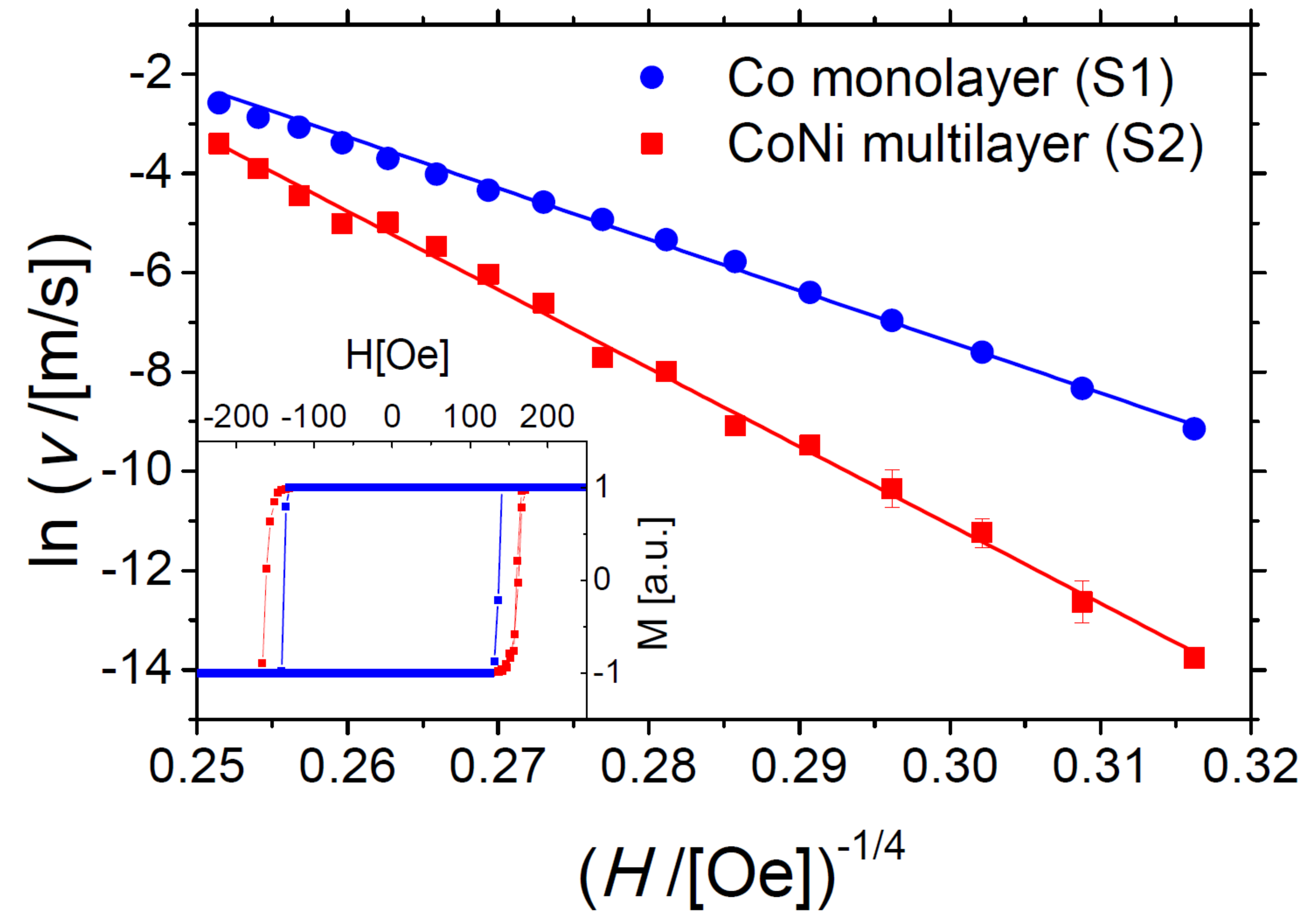}
    \caption{(Color online) Dependence of velocity on applied magnetic field for samples S1 (blue circles) and S2 (red squares). The linear behavior of the creep plot, $\ln v$ against $H^{-1/4}$, indicates that both systems are in the creep regime. The inset shows magnetic hysteresis loops for each sample.}
   \label{fig:velocity}
\end{figure}

Local magnetization hysteresis loops were obtained by PMOKE microscopy in both samples by applying successive 50 ms magnetic field pulses of increasing (decreasing) amplitudes up to ±250 Oe. The inset of Fig. \ref{fig:velocity} shows the out-of-plane hysteresis loops for each sample, whose squared shape is typical of systems with PMA. The main panel of Fig. \ref{fig:velocity} shows DW velocities corresponding to samples S1 (blue circles) and S2 (red squares) in a creep plot $\ln v$ as a function of $H^{-1/4}$. The observed linear relationship confirms that the DW dynamics, in both samples, remains in the creep regime, with the expected dynamic exponent $\mu = 1/4$~\cite{Nattermann90, Lemerle98, Chauve2000, Metaxas2007, Ferrero2013}. Domains nucleate at similar fields in both samples, but the DW velocities are higher in sample S1 which can be correlated with a sharper hysteresis loop and a smaller coercive field~\cite{Cintia2018}. 

The experimental roughness function $B(r)$, defined in Eq. \eqref{FunBr}, has been computed for DWs driven by magnetic fields of 100 Oe for S1 and 150 Oe for S2, which corresponds to approximately the same velocity in Fig.~\ref{fig:velocity}. The corresponding $\zeta$ and $B_{0}$ values were estimated from the linear fit of $\log B(r)$ as a function of $\log r$, as in the example shown in Fig. \ref{fig:rugosidad}. However, a statistical analysis is needed in order to obtain reliable roughness parameters~\cite{Guyonnet-thesis}. A more detailed discussion about this issue is included in a forthcoming article \cite{papercabrugosidad}. The resulting roughness exponents, statistically averaged over $10$ realizations, are $\zeta _{DC} = 0.73 \pm 0.04$ and $\zeta _{DC} = 0.64 \pm 0.05$ for sample S1 and S2, respectively. In both cases, roughness exponent values are close to the predicted equilibrium value $\zeta = 2/3$. In addition, the obtained values for the roughness amplitudes are $B_0 = (0.032 \pm 0.006) \, \mu \mathrm{m}^2$, for S1, and  $B_0 = (0.035 \pm 0.007) \, \mu \mathrm{m}^2$, for S2, resulting in similar DC roughness amplitudes for both samples.

\section{AC DRIVEN EVOLUTION}\label{sec:AC}

As mentioned in the Introduction, according to Nattermann \textit{et al.}, the scenario that describes the DW dynamics under the application of alternating fields refers to different frequency and instantaneous magnetic field dependent regimes. This response may be strongly non-linear and hysteretic, but after each AC cycle the system is expected to return to the initial condition. In that sense, all the cycles are similar and described by a stationary behavior, because the initial condition does not play any role. Instead, our experimental results, as presented in the following, show that the actual situation can be much more complex.

\begin{figure}[ht!]
  \centering
    \includegraphics[width=\linewidth]{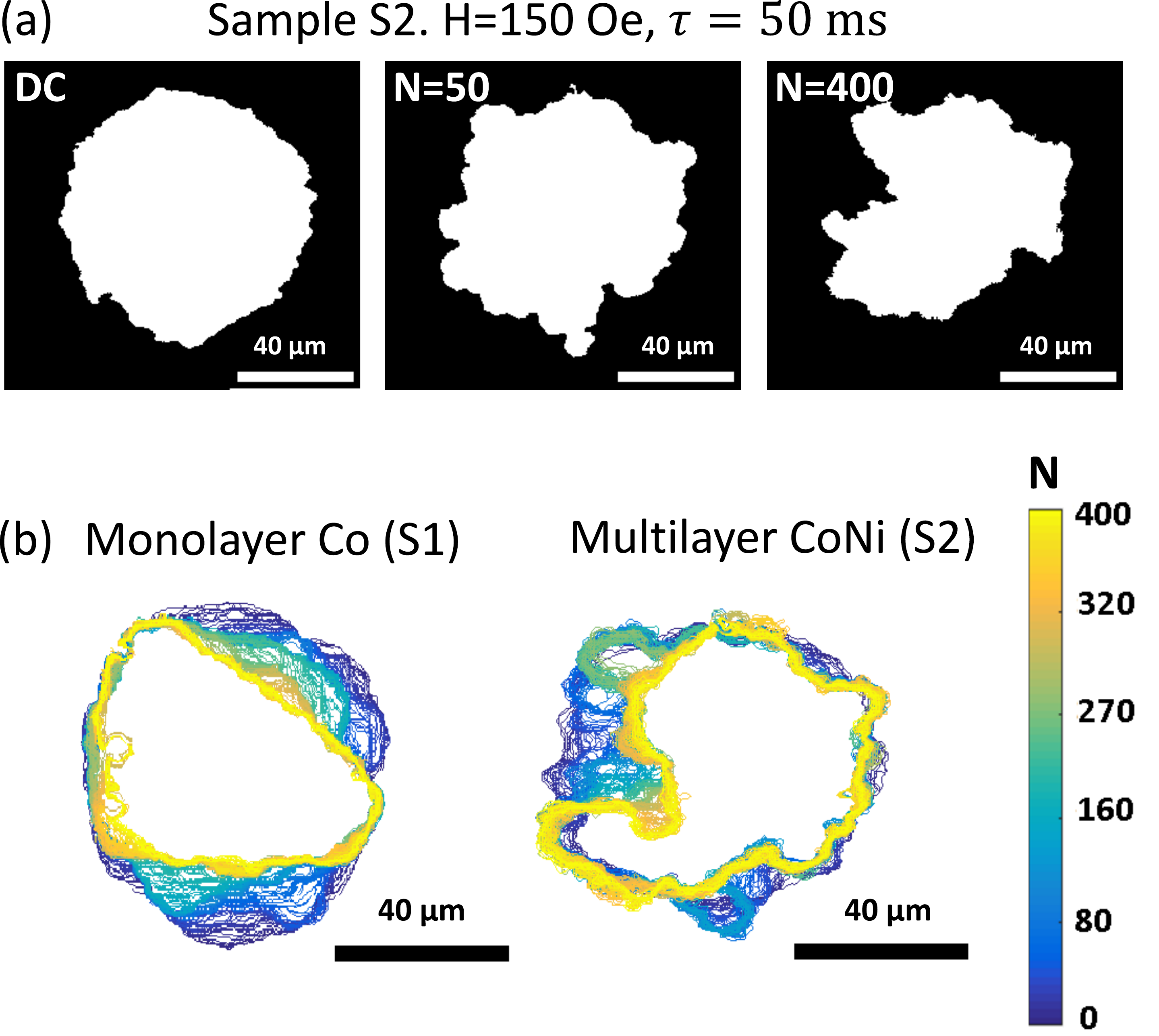}
    \caption{(Color online) (a) Examples of images taken during the AC evolution for sample S2, for different number of AC cycles, indicated in the top left corner.  (b) Superimposition of the domain's contours as a function of the number of AC cycles, from the DC growth domain (blue) to 400 AC cycles (yellow).}
    \label{fig:DomainAC}
\end{figure}

Figure \ref{fig:DomainAC} shows a typical evolution of magnetic domains during the application of an AC protocol. A particular case for sample S2 is depicted in Fig. \ref{fig:DomainAC}(a), where images of the initially DC grown domain and the same domain after applying 50 and 400 AC cycles are shown. Despite the observed DW roughness, following the DC protocol the domain grows preserving an approximately circular shape [Fig.~\ref{fig:DomainAC}(a), left]. However, it is clearly observed that after applying successive AC cycles, the shape of the magnetic domain becomes much more distorted [Fig.~\ref{fig:DomainAC}(a), center and right]. Figure \ref{fig:DomainAC}(b) shows a more detailed measurement of the evolution along 400 AC cycles for both samples. The corresponding videos are available in the Supplementary Material.

Besides the statistical character of the evolution, we found that the average characteristics of domains morphology are strongly dependent on three distinct factors: the initial conditions, the number of AC cycles in each burst and the average displacement of the DW in each half-cycle $\Delta u$. Therefore, to be able to compare and quantify observables that characterize the AC evolution, we have established the following protocol: After nucleation, all domains are grown to the same area (arbitrarily chosen to be $ \sim 4700 \,\mu \mathrm{m}^2$) by applying DC pulses with a given magnetic field and pulse duration. Then, AC magnetic field pulses with $\tau = 50 \,\mathrm{ms}$ are applied. To make the experiments for different samples comparable, the amplitude of the magnetic field pulses, $H = 100\,\mathrm{Oe}$ and $H = 150\,\mathrm{Oe}$ for S1 and S2 respectively, were chosen so that the DW velocities in each half-cycle produce an average displacement of $ \Delta u \sim 6 \, \mu \mathrm{m}$.

Direct observation of the images during the AC evolution clearly shows that the domains not only change their shape, but also loose area, as observed in Fig. \ref{fig:DomainAC}(b).  We discarded the effect of any possible remanent small DC magnetic field in the evolution of the magnetization, by analyzing the images that result from inverting the magnetic field and AC signal polarities. By observing the evolution in more detail, it can be seen that the loss of area is smooth in general, but sometimes happens abruptly, accompanied by a collapse of a portion of the domain. These abrupt jumps in area are random in time, and the averaged evolution shows a rather smoothly decreasing area.

\begin{figure}[ht!]
  \centering
    \includegraphics[width=\linewidth]{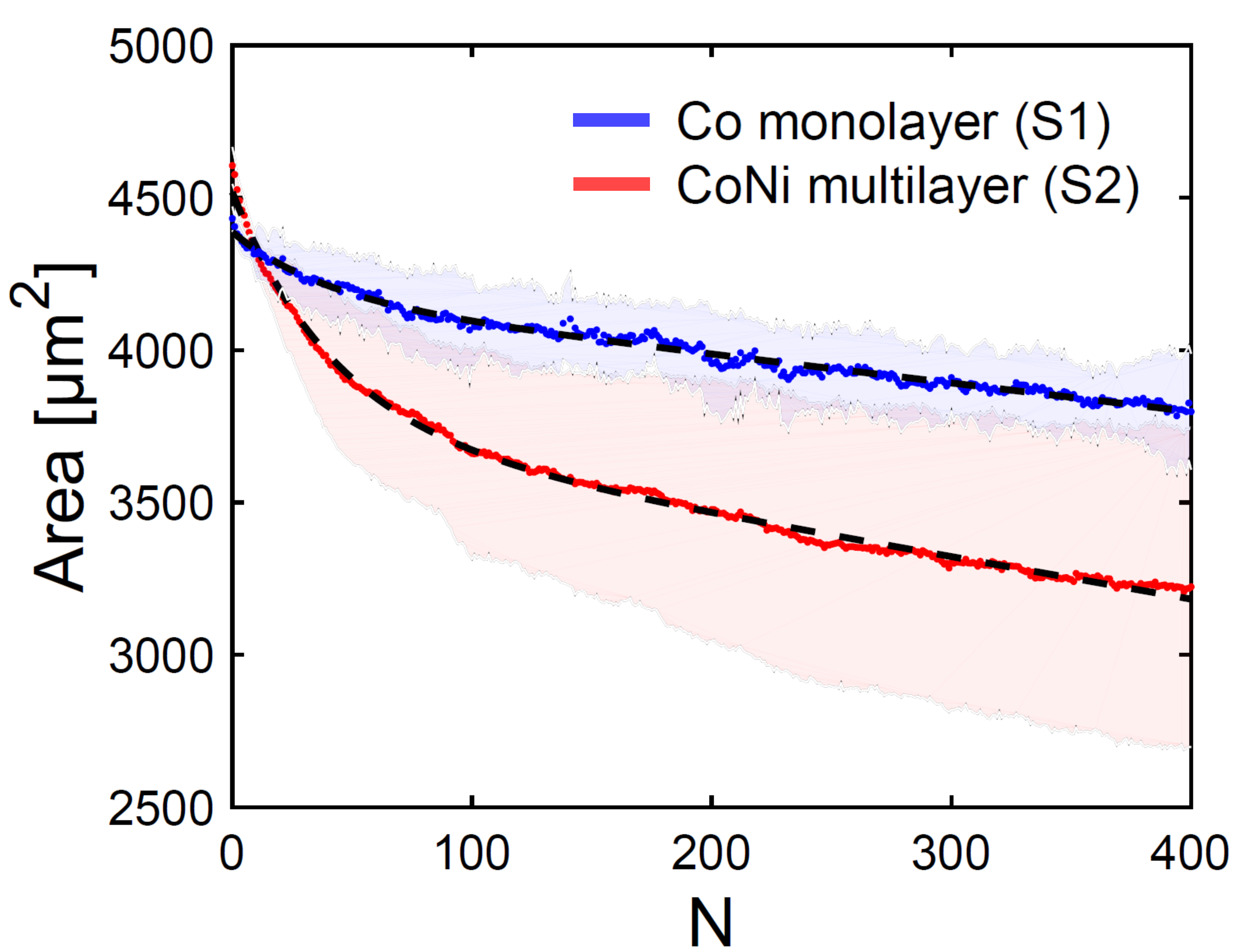}
    \caption{(Color online) Domains area as a function of the number of pulses for sample S1 (red) and sample S2 (blue). The average experimental curves were fitted with a linear combination of an exponential decay function and a linearly decreasing function (black dashed line). }
    \label{fig:AC-Area}
\end{figure}

Figure \ref{fig:AC-Area} shows the average area evolution measured in both samples, over 10 realizations of 400 AC cycles, i.e. over the repetition of the experiment 10 times. The average area of both samples decreases as $N$ increases, effect that is more noticeable in the Co monolayer. A detailed analysis of the curves shows that there is a fast decay in a characteristic AC time corresponding to $N_{AC} \sim 40$ AC cycles, followed by a slow decay (approximately linear with the number of cycles) up to the higher measured number of cycles ($N = 400$). Dashed lines in Fig. \ref{fig:AC-Area} present fits using a linear combination of exponential and linearly decreasing functions.

By observing in more detail the domains shape evolution, it can be seen that there are two scales of fluctuations. A small-scale that is normally characterized by the roughness function, and a large-scale deformation whose description goes beyond $B(r)$. Estimation of the roughness function $B(r)$ [Eq. \eqref{FunBr}] becomes a non-trivial task in strongly deformed domains, because large fluctuations in $u(\theta)$ give rise to overhangs in $u(z)$ and makes the computation of the $B(r)$ function not straighforward. An estimation of the AC roughness exponent in sample S1 has been done, by averaging the resulting exponent of 10 domains obtained after applying 400 AC cycles. The obtained AC roughness exponent for S1, $\zeta _{AC}$, resulted slightly larger, though indistinguishable within the experimental uncertainty, than the original DC exponent, $\zeta _{DC}$  (see Table \ref{Tabla:Rug}). For the roughness amplitude for sample S1 we found that ${B_0}_{AC} > {B_0}_{DC}$ (see Table \ref{Tabla:Rug}), which is representative of the observed larger fluctuations after the application of the AC protocol. For sample S2 the strongly distorted domains shape prevented a clear estimation of the roughness exponent and roughness amplitude after 400 AC cycles.

\begin{table}[ht!]
\centering
\begin{tabular}{|c|c|c|c|c|}
\hline
Sample & $\zeta_{DC}$ & $\zeta_{AC}$ & ${B_0}_{DC}$($\mu $m$^{2}$)  & ${B_0}_{AC}$($\mu $m$^{2}$) \\ \hline
S1     & $0.73 \pm 0.04$                   & $0.79 \pm 0.03$                   & $0.032 \pm 0.006$        & $0.08 \pm 0.03$        \\ \hline
S2      &  $0.64 \pm 0.05$                  & -                           & $0.035 \pm 0.007$        & -              \\ \hline
\end{tabular}
\caption{Comparison between roughness parameters in samples S1 and S2 after DC growing and after the application of 400 AC cycles (only for the S1 sample). The applied field is $100 \, \mathrm{Oe}$ for S1, and $150 \, \mathrm{Oe}$ for S2.
}
\label{Tabla:Rug}
\end{table}

\begin{figure}[t!]
  \centering
    \includegraphics[width=\linewidth]{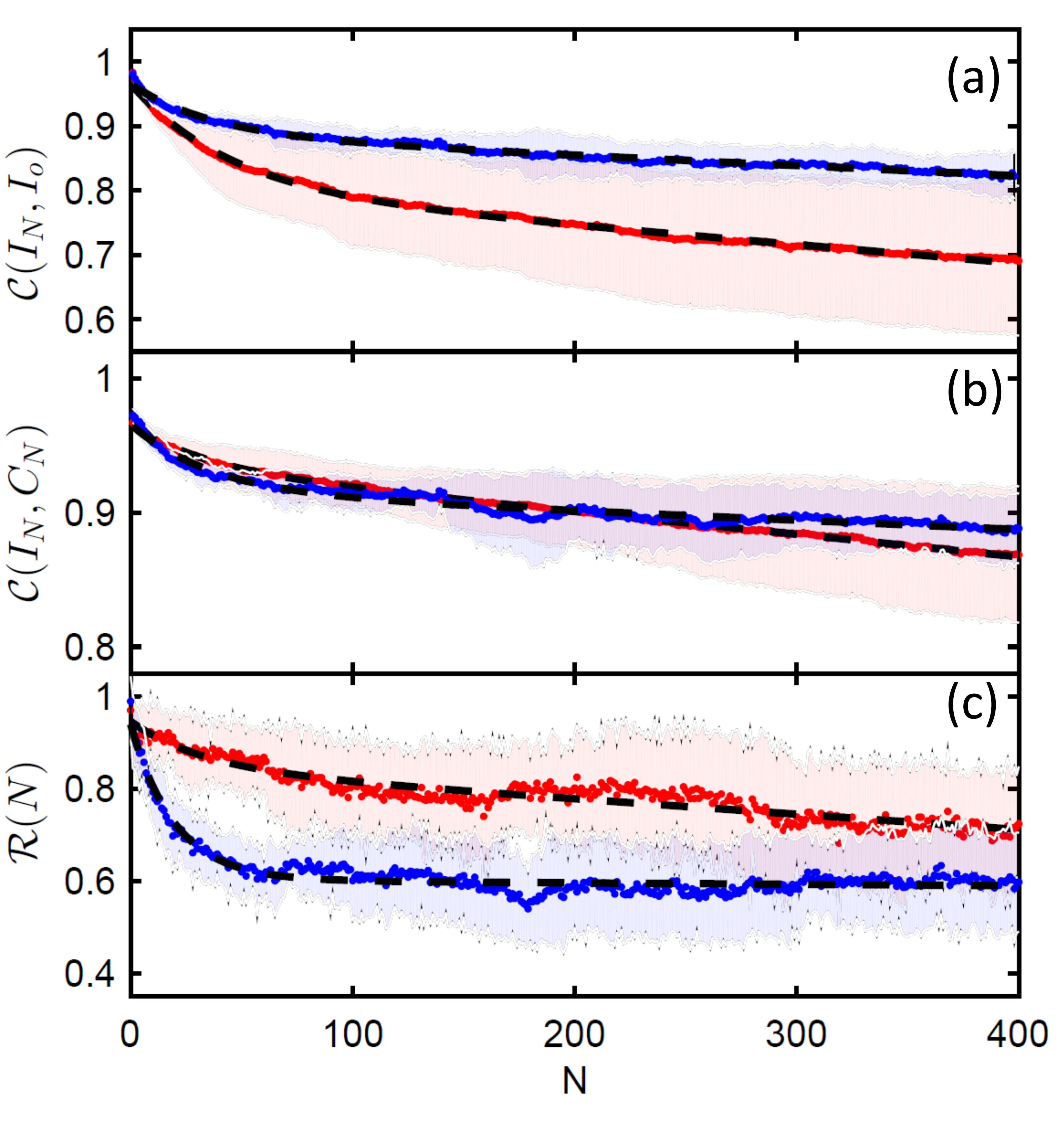}
    \caption{(Color online) Evolution of different observables as a function of the number of cycles for sample S1 (red) and sample S2 (blue). Average and variance over 10 realizations are shown.  Normalized correlation [see Eq. \eqref{FunCorr}] with the initial domain ${\cal C}(I_N,I_0)$ (a) and with a circular domain of the same area ${\cal C}(I_N,C_N)$ (b), and ratio between actual area and the area of a circle of equal perimeter ${\cal R}(N)$(c). In all cases the average curves were fitted with a linear combination of an exponential decay function and a linear decreasing function (black dashed line).}
    \label{fig:AC-Corr}
\end{figure}

In order to evidence the large-scale deformation, in the following we analyze the evolution of different observables as a function of the number of applied AC pulses. With the aim of characterizing the evolution of the domain from its initial condition, we computed the correlation between the images after applying $N$ AC cycles $I_N (x,y)$ and the image corresponding to the initial condition $I_0 (x,y)$,  ${\cal C} (I_N,I_0)$, as defined in Eq. \eqref{FunCorr}. The results displayed in Fig. \ref{fig:AC-Corr}(a) show that there is a fast decay of the correlation in a characteristic number of AC cycles, $N_{AC} \sim 30-40$, followed by a slow decay (approximately linear with the number of cycles). The value of $N_{AC}$ is of the same order as the one found for the decay of the area in Fig. \ref{fig:AC-Area}. The decay in this correlation takes into account the modification in both the shape and the area of the domains.   As can be observed in Fig. \ref{fig:DomainAC}, original domains are approximately circular.
Therefore, to separately characterize area loss and deformation, we further analyze the evolution of two observables: (\textit{i}) The correlation between the images after applying $N$ AC pulses $I_N (x,y)$ and a perfect circular domain with the same area $A(N)$ as the domain at cycle $N$, ${\cal C}(I_N,C_N)$, and (\textit{ii}) the ratio between the area $A(N)$ after $N$ cycles and the area of a perfect circle with exactly the same perimeter $P(N)$ of the domain after $N$ cycles, ${\cal R}(N) = A(N)/A_P(N) = 4 \pi A(N)/P(N)^2$. The corresponding evolutions are shown in Figs. \ref{fig:AC-Corr}(b) and \ref{fig:AC-Corr}(c) respectively.  The evolution is again well described by a fast decay before a characteristic AC number of cycles $N_{AC} \sim 30$, followed by a slower linear decay.  Noticeably, the evolution of ${\cal R}(N)$ for sample S2 displays a sharper initial decrease and a very slow decay with an almost constant value up to $N = 400$.

\begin{figure}[t!]
  \centering
    \includegraphics[width=\linewidth]{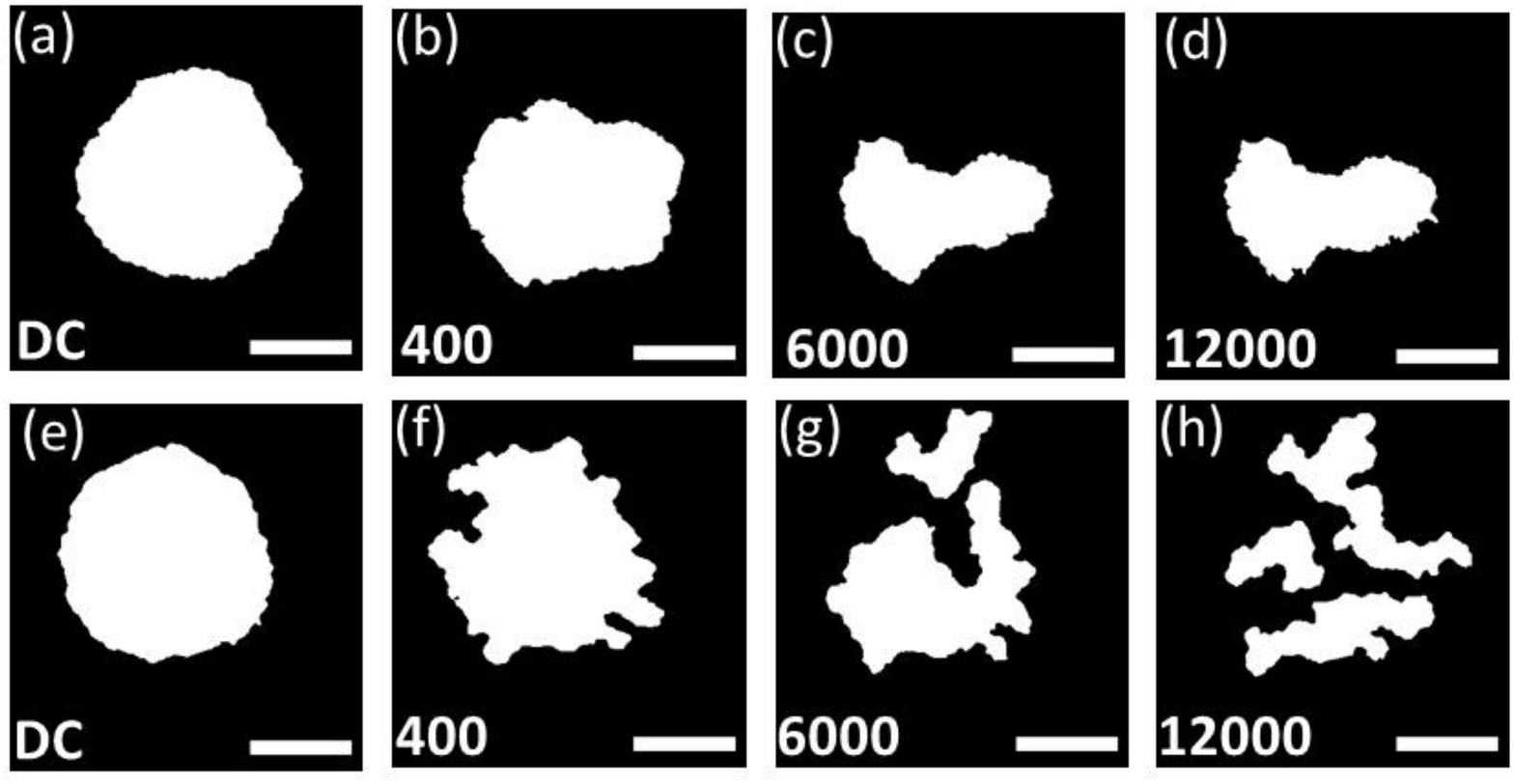}
    \caption{Example of domains evolution after applying a large number of cycles  for sample S1 (a-d) and sample S2 (e-h). Scale bars are $40  \,\mu \mathrm{m}$.}
    \label{fig:AC-large}
\end{figure}

The evolution of all the observables shows a crossover from a rapid decay to a slow linear decay
in a transient number of cycles (around $N_{AC} = 30-40$ for this particular set of experimental conditions). The tendency measured up to $N = 400$ AC cycles suggests a final stable magnetized state (nearly saturation).
A study up to much larger number of cycles, statistically representative, should involve many technical challenges as for example the statistical repetitivity or the mechanical stability of the PMOKE microscope. Still, images taken after applying a large number of AC cycles $N$ (up to 12000) show that domains continue to evolve, with very different behaviors on each sample. Typical images for each sample, representative of different stages, are shown in Fig. \ref{fig:AC-large}. For sample S1, there is a continuous decrease of the domain area, with an eventual full collapse. On the other hand, domains in sample S2 break into several sub-domains, and no saturation is observed up to the largest times involved in our experiments.

\section{Discussion and Conclusions}\label{sec:Conclusions}

The discussion about the underlying physics behind the AC evolution of domains and DWs is not obvious, and deserves further investigation. The model proposed in Refs.~~\cite{Lyuksyutov1999, Nattermann2001, Glatz2003, Nattermann2004} traces an analogy between DWs and one-dimensional elastic interfaces in the presence of quenched random disorder. Although it captures the mean ingredients of the AC dynamics in terms of the DC counterpart, it is unable to predict the AC driven behavior observed in the present work. This model is based on a free energy quasi-equilibrium description, assuming a random week disorder distribution and neglecting long range dipolar interactions. Furthermore, this model also assumes a flat DW configuration, disregarding DW fluctuations and its dynamics.

Numerical simulations and experiments on ferroelectric DWs show that increasing dipolar interactions might lead to deformation of domains and rougher DWs~\cite{Guyonnet-thesis}. In order to evaluate the relevance of dipolar interaction in our samples, typical energy and length scales, computed from micromagnetic parameters are presented in Table \ref{Tabla-param}.

A smaller $Q$ factor indicates a larger dipolar influence. Contrary to what could be intuitively expected~\cite{Guyonnet-thesis}, the Co monolayer presents less distorted domains after application of the AC protocol but has a smaller $Q$ factor than the CoNi multilayer. Furthermore, by comparing the exchange length $l_{ex}$, different values are obtained for each sample, with a larger value (and thus smaller dipolar effect) obtained for CoNi multilayers. In addition, estimations of the DW energy density and DW width give very similar values for both samples. All these comparisons point to a smaller influence of the dipolar interaction in sample S2, which presents however more distorted domains during the AC driven evolution. Therefore, at least after a first qualitative analysis, dipolar interaction does not seem to be the key responsible of domains deformation.

\begin{table}[t!]
\centering
\footnotesize
\begin{tabular}{|c|c|c|c|c|}
\hline
Parameter                                                            & Formula                                                 & Units & Pt/Co/Pt & \begin{tabular}[c]{@{}c@{}}CoNi\\ multilayer\end{tabular} \\ \hline
\begin{tabular}[c]{@{}c@{}}Anisotropy\\   constant\end{tabular}      & $K$                                                       & kJ/m$^3$ & 364               & 340              \\ \hline
\begin{tabular}[c]{@{}c@{}}Saturation\\   magnetization\end{tabular} & $M_S$                                                    & kA/m  & 910               & 540              \\ \hline
\begin{tabular}[c]{@{}c@{}}Exchange\\ constant\end{tabular}          & $A$                                                       & pJ/m  & 14                & 15               \\ \hline
\begin{tabular}[c]{@{}c@{}}Dipolar energy \\ scale\end{tabular}      & $K_d = \mu _0 M_S ^2 /2$ & kJ/m$^3$ & 520               & 183              \\ \hline
\begin{tabular}[c]{@{}c@{}}DW energy \\ density\end{tabular}         & $\sigma = \sqrt{AK}$      & mJ/m$^2$ & 2.26              & 2.26             \\ \hline
\begin{tabular}[c]{@{}c@{}} Q-factor                                                        \end{tabular}     & $Q=K/K_d$                                                & -     & 0.7              & 1.86              \\ \hline
\begin{tabular}[c]{@{}c@{}}DW \\ width\end{tabular}                  & $\Delta = \sqrt{A/K}$     & nm    & 6.2               & 6.6              \\ \hline

\begin{tabular}[c]{@{}c@{}}Exchange\\ length\end{tabular}      & $l_{ex}=\sqrt{A/K_d}$                          & nm    & 0.16                & 0.27                \\ \hline
\end{tabular}
\caption{Typical micromagnetic parameters and energy and length scales for each sample. Micromagnetic parameters (anisotropy constant $K$, saturation magnetization $M_S$ and exchange constant $A$) were extracted from Refs. \onlinecite{Metaxas2007} and \onlinecite{Rojas2016} for Pt/Co/Pt and CoNi multilayers, respectively. Energy and length scales were estimated from given micromagnetic parameters.}
\label{Tabla-param}
\end{table}

Although it has been shown that the DMI might play an important role in the studied samples, since in the present  experiments we are not applying in-plane fields, the asymmetric response of DWs due to the DMI is not unveiled. However, the DW width and DW energy density do depend on the DMI and thus it is effectively influencing  the resulting dynamics.

A careful inspection of the obtained images suggests that an enhancement of the influence of strong pinning centers with respect to the DC evolution could be one of the key ingredients in the AC dynamics of DWs. Plausibly, the fact that during AC experiments the DW moves back and forth across the disorder energy landscape an increasing number of times, causes the DW to further explore the pinning centers and hence to progressively reach a more pinned configuration. How the DW is able to explore the energy landscape should be related to the time scale associated to the collective response of the magnetic moments on the DW. Therefore, the DW response would depend on both the rise and fall time of the AC field and its switching rate.
Preliminary experiments indicate that the AC driven evolution is very dependent on the switching rate, but a systematic study is necessary to be conclusive about its role in the deformation of domains.

In summary, AC driven DW evolution has been investigated by means of PMOKE microscopy, in two different ultra-thin magnetic films with PMA.  The DC characterization is well predicted by the elastic interface model, and similar dynamic exponents are obtained in both samples.

However, striking unexpected features are observed under the application of alternating square pulses: a strong DW deformation at scale larger than that involved in the interface roughness, together with a decrease of the total domain area are observed.  Whereas the roughness exponent is not the best parameter to describe the full domain deformation, the increase in the roughness amplitude is in qualitative agreement with the observations. We identified alternative observables that are able to describe the large scale AC evolution even for the case of highly distorted domains, from which we determined a typical transient AC number of cycles, around $N_{AC} = 30-40$ for the protocol used in our experiments.

The evolution at very large number of AC cyles displays very different features in both samples: whereas there is a tendency to domain collapse for Pt/Co/Pt, domains break up for CoNi multilayers.  Both states seem to be robust attractors for the dynamics trayectories, but extensive and statistically representative studies for extremely large number of AC cycles are necessary to extract rigorous conclusions.

Numerical simulations as well as systematic complementary experimental studies are also necessary to find out the main physical ingredient responsible of the observed behavior. A major influence of strong pinning centers and/or the magnetic field switching time and switching rate in each AC cycle, are the main candidates.
Overall, we think that the reported experiments open a rich and interesting variety of future work in the field of domain wall dynamics.

\textbf{Acknowledgements}: The authors acknowledge illuminating discussions with A. B. Kolton. We would like to thank V. Bekeris, C. Gourdon and M. Tortarolo for their aid during the design and tunning of the MO setup. This work was partially supported by the National Scientific and Technical Research Council - Argentina (CONICET), the University of Buenos Aires, the University of Cuyo and the ANPCyT.

\bibliography{refs_MO}

\end{document}